# Beyond Pairwise Distance: Cognitive Traversal Distance as a Holistic Measure of Scientific Novelty

Yi Xiang[1,2], Pascal Welke[2,3], Chengzhi Zhang[1], Jian Wang[2,*]
1. Nanjing University of Science and Technology, 210094, Nanjing, China
2. Lancaster University Leipzig, 04109, Leipzig, Germany
3. TU Wien, 1040, Vienna, Austria

*Corresponding author

**Abstract:**

Scientific novelty is a critical construct in bibliometrics and is commonly measured by aggregating pairwise distances between the knowledge units underlying a paper. While prior work has refined how such distances are computed, less attention has been paid to how dyadic relations are aggregated to characterize novelty at the paper level. We address this limitation by introducing a network-based indicator, *Cognitive Traversal Distance* (CTD). Conceptualizing the historical literature as a weighted knowledge network, CTD is defined as the length of the shortest path required to connect all knowledge units associated with a paper. CTD provides a paper-level novelty measure that reflects the minimal structural distance needed to integrate multiple knowledge units, moving beyond mean- or quantile-based aggregation of pairwise distances. Using 27 million biomedical publications indexed by OpenAlex and Medical Subject Headings (MeSH) as standardized knowledge units, we evaluate CTD against expert-based novelty benchmarks from F1000Prime-recommended papers and Nobel Prize–winning publications. CTD consistently outperforms conventional aggregation-based indicators. We further show that MeSH-based CTD is less sensitive to novelty driven by the emergence of entirely new conceptual labels, clarifying its scope relative to recent text-based measures.

# 1. Introduction

Novelty is widely understood as the recombination of existing knowledge units in new or unusual ways. Schumpeter's classic definition of innovation as "the establishment of a new production function" through unprecedented combinations of factors (Schumpeter, 1934) laid the foundation for a combinatorial perspective that remains central to the study of scientific novelty. Building on this tradition, scholars increasingly assess novelty by examining how academic papers recombine prior knowledge, typically operationalized through the newness or rarity of pairwise combinations among existing concepts (Uzzi et al., 2013; Wang et al., 2017). These indicators aim to provide ex ante measures of novelty that are independent of downstream outcomes such as citations, impact, or disruptiveness (Funk & Owen-Smith, 2017; Wang et al., 2023).

While pairwise approaches have generated important insights, they capture only a limited aspect of the cognitive effort involved in producing new knowledge. Scientific contributions rarely hinge on a single novel combination; instead, they emerge from the integration of multiple ideas, methods, and concepts into a coherent whole (Wilson, 1999). Incorporating the global structural information of these multiple knowledge units into novelty analysis opens new avenues for understanding the relationship between knowledge composition and scientific innovation—a depth of insight that is unattainable via traditional pairwise methods. Specifically, two papers may exhibit similar distributions of pairwise distances yet differ in how their knowledge units are jointly organized.

To address this limitation, we conceptualize scientific research as a cognitive traversal of an intellectual landscape defined by the historical structure of knowledge. In this landscape, each knowledge unit occupies a position determined by its established relationships with others, and producing a scientific paper requires navigating this space to identify relevant concepts and integrate them into a coherent contribution. This navigation may follow a targeted search guided by theoretical understanding (Fleming & Sorenson, 2001, 2004; Gruber et al., 2013) or proceed through more serendipitous wandering across unfamiliar domains (March, 1991; Simonton, 1999; Rosvall & Bergstrom, 2008). Regardless of the specific trajectory, we argue that the minimal conceptual distance required to connect all knowledge units in the final product captures the cognitive effort involved and, in turn, the degree of novelty perceived by the audience of the research.

Building on this conceptualization, we propose a new paper-level novelty measure: *Cognitive Traversal Distance* (*CTD*). We model the knowledge units contained in a paper as nodes in a weighted network, where edge weights represent conceptual distances derived from historical literature. CTD is defined as the length of the shortest traversal that connects all knowledge units. This traversal represents the minimal cognitive effort required to integrate all elements into a coherent contribution.

Papers whose knowledge units are concentrated within dense, closely connected regions of the knowledge space yield shorter minimal traversals and thus exhibit lower novelty, whereas papers that span dispersed or weakly connected regions require longer traversals and exhibit higher novelty.

To evaluate CTD, we analyze 27 million biomedical papers indexed in OpenAlex, using MeSH terms as knowledge units. For each paper, we derive CTD scores using established pairwise conceptual distances calculated from the preceding five years of literature. We benchmark these scores against expert assessments using two validated datasets: Nobel Prize–winning publications (Li et al., 2019) and F1000Prime recommendations[1]. Across both settings, CTD consistently distinguishes novel papers and outperforms conventional mean- (Ruan et al., 2025) or quantile-based (Uzzi et al., 2013) aggregation metrics. We further compare CTD against recent high-performing text-based measures relying on new words and combinations (Arts et al., 2025). While CTD exhibits superior performance in predicting F1000Prime papers, text-based indicators prove most effective for Nobel Prize–winning publications. This suggests that MeSH-based traversal is less sensitive to novelty driven by the emergence of entirely new conceptual labels, highlighting a specific limitation of our measure.

This study makes three contributions. First, it introduces a conceptualization of scientific novelty grounded in the cognitive traversal of a structured knowledge landscape, moving beyond dyadic recombination toward holistic integration. Second, it develops a novel operational measure, Cognitive Traversal Distance, that captures the minimal cognitive effort required to integrate all knowledge units in a paper. Third, it provides large-scale empirical validation demonstrating that CTD is both effective and robust across datasets.

## 2. Related Work

### 2.1 Combinatorial Approaches to Novelty Assessment

Most existing methods draw on a combinatorial view, conceptualizing scientific advances as the recombination of existing knowledge units. Within this perspective, prior studies differ along three dimensions: the definition of knowledge units, pairwise distance measurement, and paper-level aggregation.

Early work relied on references as proxies for knowledge, using journals (Uzzi et al., 2013; Lee et al., 2015; Mukherjee et al., 2016; Wang et al., 2017), reference titles (Shibayama et al., 2021), or cited documents themselves (Trapido et al., 2015; Matsumoto et al., 2021) as basic units of analysis. However, references are cited for heterogeneous reasons that do not necessarily reflect knowledge absorption or conceptual integration (MacRoberts & MacRoberts, 1996; Ding et al., 2014), which complicates interpretation. To obtain more direct representations of conceptual content, later studies adopted lexical units such as keywords (Yan et al., 2020), noun phrases (Arts et al., 2025), named entities (Liu et al., 2022; Liu et al., 2024), or structured representations such as problem–method pairs (Luo et al., 2022). While these approaches improve semantic precision, their fine granularity introduces challenges related to polysemy, synonymy, and contextual ambiguity. In this study, we employ MeSH terms, an expert-curated, domain-specific vocabulary that has been widely used in novelty assessment (Boudreau et al., 2016; Mishra & Torvik, 2016; Chai & Menon, 2019; Ruan et al., 2025), to ensure conceptual consistency and interpretability.

---

[1] https://archive.connect.h1.co

Once knowledge units are defined, studies propose various measures of pairwise novelty or distance. A fundamental distinction concerns rarity versus newness: rarity-based measures emphasize how unusual or atypical a pairwise combination is relative to historical baselines (Uzzi et al., 2013; Lee et al., 2015), whereas newness-based measures focus on whether a combination has not previously occurred (Wang et al., 2017; Arts et al., 2025). Methodologically, some studies rely on first-order occurrence matrices (paper × knowledge unit) to compute observed-to-expected ratios, z-scores, or cosine similarity measures (Uzzi et al., 2013; Lee et al., 2015; Mishra & Torvik, 2016). Others extend to second-order co-occurrence matrices (knowledge unit × knowledge unit) (Wang et al., 2017) or continuous representations derived from embedding models (Liu et al., 2022; Yin et al., 2023).

To obtain a paper-level novelty score, prior work aggregates pairwise distances using summary statistics such as quantiles (e.g., the 10th percentile; Uzzi et al., 2013; Lee et al., 2015; Mukherjee et al., 2016; Yin et al., 2023), averages (Ruan et al., 2025), or sums (Wang et al., 2017; Shibayama et al., 2021). Other approaches quantify the number of new or highly novel pairs (Arts et al., 2025) or the proportion of newly introduced combinations (Liu et al., 2022).

Despite their methodological diversity, these approaches share a key limitation: novelty is assessed exclusively at the dyadic level between knowledge units. Paper-level scores are then obtained through set aggregation without modelling how all units jointly relate to one another. As a result, these measures cannot distinguish between papers with similar pairwise distances but substantially different structural configurations.

## 2.2 Toward a Holistic Model of Knowledge Integration

A small but growing body of work has begun to move beyond pairwise combinations of knowledge integration. Notably, Shi and Evans (2023) model all cited journals of a paper as nodes connected by a hyperedge and estimate the joint probability of their co-occurrence using a hypergraph framework. This approach represents an important step toward capturing higher-order combinatorial structure. However, it does not explicitly model how the relational configuration or structural distances among knowledge units shape the cognitive effort required to integrate them, nor does it provide an interpretable distance-based measure of novelty.

## 2.3 Alternative Approaches Beyond Combinatorial Perspectives

Other research estimates scientific novelty using non-combinatorial approaches. Some measure semantic distance between a focal paper and its nearest neighbors in document space (Jeon et al., 2023; D'Aniello et al., 2024). Others apply supervised or unsupervised machine-learning models to predict novelty or impact (Wang et al., 2024; Wu et al., 2025), or leverage large language models to assess scientific creativity or originality (Tan et al., 2025; Huang et al., 2025). While promising, these approaches often lack a clear theoretical mapping between model outputs and cognitive mechanisms of novelty and offer limited interpretability regarding how specific knowledge units contribute to the assessment.

## 3. Our Approach: Cognitive Traversal Distance (CTD)

Measuring scientific novelty requires more than identifying unusual pairwise combinations of knowledge units. Conceptually, producing a research paper involves integrating multiple ideas, concepts, or methods into a coherent contribution. We argue that this integration process can be conceptualized as a cognitive traversal across a structured knowledge landscape in which scientific concepts reside. Building on this intuition, we propose *Cognitive Traversal Distance (CTD)*, a holistic novelty measure that captures the minimal cognitive effort required to connect all knowledge units in a paper.

Scholars have long conceptualized knowledge creation as a search process through a structured knowledge space (Ward et al., 1997). This search may follow a targeted trajectory guided by scientific understanding or analogical reasoning that functions as a "map," directing effort toward promising regions and pruning unproductive paths (Fleming & Sorenson, 2001; 2004; Gruber et al., 2013). Alternatively, it may proceed through more serendipitous processes, akin to random walks or associative wandering across loosely connected ideas (March, 1991; Simonton, 1999; Rosvall & Bergstrom, 2008). Across these perspectives, novelty arises from the intellectual effort required to traverse distance in knowledge space (Ward et al., 1997).

Prior bibliometric approaches primarily operationalize novelty through new or unusual pairwise combinations of knowledge units (Uzzi et al., 2013; Lee et al., 2015; Wang et al., 2017). While influential, these measures remain fundamentally dyadic and therefore overlook higher-order structure. Novelty may stem from a "long search path," that is, a sequence of incremental steps required to connect fragmented regions of knowledge (Kneeland et al., 2020). Such traversal-based novelty is not adequately captured by existing pairwise metrics.

## 3.1 Conceptualizing Novelty as Cognitive Traversal Distance

To address this limitation, we draw a parallel between the knowledge search process underlying scientific research and classic traversal problems in graph theory, i.e., the traveling salesperson problem. We conceptualize the novelty of a scientific paper as its *Cognitive Traversal Distance* (*CTD*), defined as the length of the shortest path required to visit all knowledge units of the paper at least once, given the historical structure of the knowledge space.

A central insight of this approach is that novelty depends not merely on isolated distances between pairs of concepts, but on the global configuration of all knowledge units involved. CTD captures both the presence of distant or unusual conceptual connections, and the cumulative effort required to traverse a sequence of intermediate knowledge units that bridge these distances. A shorter CTD indicates that a paper's knowledge units are clustered within a compact region of the knowledge landscape and therefore require limited cognitive effort to integrate. A longer CTD suggests that the paper spans multiple distant regions and embodies greater conceptual novelty.

Importantly, CTD does not aim to reconstruct the authors' actual cognitive search process. Authors may explore many ideas that do not appear in the final publication, take circuitous routes despite the existence of shorter conceptual connections. Instead, we intentionally focus on the perspective of the scientific audience, that is, how peer scientists evaluate a published contribution relative to the existing structure of knowledge. In this sense, CTD reflects the minimal cognitive traversal implied by the final product, rather than the realized path taken during research. This perspective aligns with the creativity literature's emphasis that novelty and creativity are assessed in the eye of the audience (Amabile, 1983).

## 3.2 Operationalization and Comparison

To operationalize CTD, we represent the set of knowledge units contained in a paper (proxied by MeSH terms) as nodes in a fully connected, weighted, undirected network. Each edge weight reflects the conceptual distance between a pair of knowledge units, computed using established pairwise novelty formulations (details are provided in the Data and Methods section). This network encodes the historical structure of the knowledge space at the time of publication.

CTD is then defined as the length of the shortest traversal that visits all nodes at least once, given these pairwise distances. By construction, this measure incorporates the entire configuration of knowledge units, rather than aggregating independent dyadic distances.

Table 1: Cognitive Traversal Distance (CTD) Compared with Pairwise Novelty Measures

|  | Distance (Paper 1) | Distance (Paper 2) |
|---|---|---|
| AB | 1 | 3 |
| AC | 4 | 3 |
| AD | 5 | 3 |
| BC | 3 | 3 |
| BD | 4 | 3 |
| CD | 1 | 3 |
| Pairwise Distance: Average | 3 | 3 |
| Pairwise Distance: Maximum (or 90$^{th}$ percentile) | 5 | 3 |
| Cognitive Traversal Distance | 5 (A-B-C-D) | 9 (A-B-C-D) |

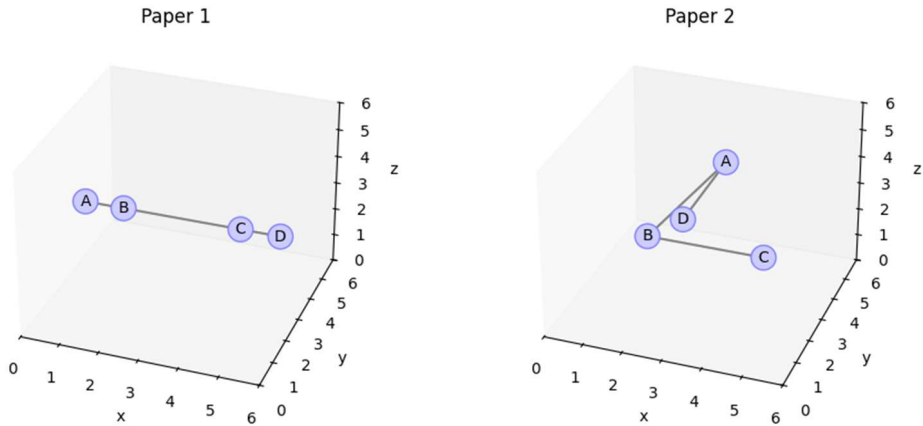

**Figure 1: Illustration of Cognitive Traversal Distance (CTD).** Network positions of four knowledge units embedded in a three-dimensional historical knowledge space for two illustrative papers. Nodes represent knowledge units, and edges depict the shortest traversal connecting all elements. Although both papers share similar pairwise distance statistics, their minimal traversals differ in length. CTD captures this global structural difference by measuring the total distance of the shortest traversal required to integrate all knowledge units, highlighting novelty that is not detectable using pairwise aggregation alone.

**Table 1** presents two illustrative examples, Paper 1 and Paper 2, each involving four knowledge units. Although the two papers exhibit identical average pairwise distances, their global structures differ substantially. **Figure 1** visualizes the four knowledge units embedded in a three-dimensional historical knowledge space and traces the shortest traversal connecting all components. Under CTD, Paper 2 is more novel than Paper 1 because connecting its knowledge units requires a longer and more effortful traversal. In contrast, the average pairwise distance metric assigns the same level of novelty to both papers, failing to capture higher-order structural differences. The maximum (or 90th percentile) pairwise distance measure would even rank Paper 1 as more novel, as it is dominated by a single extreme dyadic distance (between A and D), while ignoring the presence of intermediate concepts (B and C) that reduce the overall integration effort. Taken together, this example illustrates how CTD captures an important and previously overlooked dimension of scientific novelty: the global cognitive effort required to integrate multiple knowledge units into a coherent contribution.

## 4. Methods and Data

A practical implementation of our Cognitive Traversal Distance depends on specific choices of knowledge units and distances between them. In this chapter, we use Medical Subject Headings as knowledge units and measure dyadic distances based on historical linkage relations between knowledge units, on which we present the computation and validation of the CTD measure.

### 4.1 Data Collection

This study conducts an empirical analysis in the biomedical domain. A key advantage of this domain is that the U.S. National Institutes of Health (NIH) assigns MeSH to biomedical publications. MeSH terms comprehensively describe the knowledge content of a paper, including research objects, methods, and conceptual focus, and thus provide well-defined proxies for knowledge units in novelty measurement.

To collect the experimental data, we download the OpenAlex snapshot[2] dated March 24, 2025. The snapshot provides records for scientific papers containing multiple identifiers (OpenAlex ID, PubMed ID, DOI), the associated MeSH terms, and metadata including citation counts, author information, affiliations, and countries. We exclude records with missing PubMed IDs and restrict the dataset to publications from 1900 to 2020, yielding an initial sample of 31,849,996 papers. Because our novelty measure is defined over combinations of knowledge units, we exclude records with empty MeSH-term lists (4,347,263 papers) and those containing only a single MeSH term (79,284 papers). The final analytical dataset consists of 27,423,449 papers.

### 4.2 Paper-Level Novelty Measure: Cognitive Traversal Distance

We characterize relationships among knowledge units using a network-based representation, illustrated in **Figure 2**. For each focal paper, we first construct a historical knowledge network based on all papers published in the preceding five years, where nodes correspond to observed MeSH terms and edges capture their co-occurrences in papers of the past five years.

Subsequently, we construct a sub-network specifically for the focal paper, comprising all MeSH terms appearing within it. The edge weights are derived from the historical network, using five distinct calculation methods detailed in Section 4.3.

We then compute CTD as the length of the shortest traversal that connects all knowledge units in the focal paper. Formally, CTD corresponds to the minimum total weight of a walk that visits all nodes in the focal paper's knowledge network at least once. Computing this shortest traversal is a variant of the Traveling Salesperson Problem based on the distance matrix of the focal paper and computationally hard

---

[2] https://openalex.s3.amazonaws.com/browse.html#data/

(Garey & Johnson, 1972). However, practical solutions exist which can quickly solve instances at the scale of hundreds of knowledge units. We use Google OR-Tools[3], a high-performance optimization library for routing and combinatorial optimization, to quickly compute the shortest traversal for each focal paper. As the average number of MeSH terms in focal papers is around 10 (as depicted in **Figure 4**), this approach is viable.

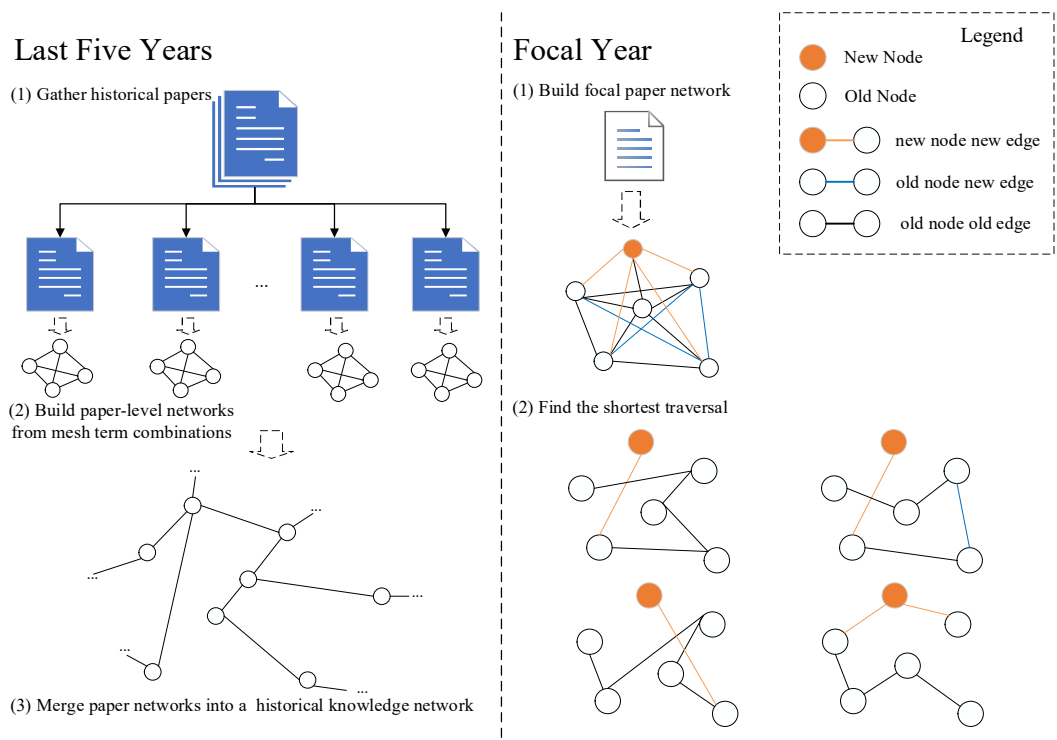

**Figure 1. Construction of Historical and Focal Knowledge Networks and Computation of Cognitive Traversal Distance.** Left panel: Papers published in the five years preceding the focal year are collected, and each paper is represented as a fully connected network of MeSH terms. These paper-level networks are then merged to form a historical knowledge network capturing prior conceptual relationships. Right panel: For a focal paper, MeSH terms are represented as nodes and pairwise distances as edge weights, with nodes and edges classified as new or old relative to the historical network. CTD is defined as the length of the shortest traversal (walk) that visits all knowledge units at least once, representing the minimal cognitive effort required to integrate the paper's knowledge units given the historical structure of knowledge.

## 4.3 Pairwise Distance Measures

In the focal paper's subgraph (**Figure 2** right panel), it may happen that some MeSH terms in the focal paper do not appear in the historical graph. In this case, we call them *new*, otherwise *old*. Correspondingly, a pair of nodes in the focal paper graph can fall into three categories: (1) *new node new edge:* the pair contains one or two new nodes, (2) *old node new edge:* the pair contains two old nodes that are not connected in the historical graph, and (3) *old node old edge:* an edge in the historical graph connecting two old nodes. Methods that operate only on existing edges or first-order occurrence matrices fail to account for all new pairs in category (1) and (2). To systematically compare alternative approaches and to ensure coverage of all edge types, we implement five methods for pairwise novelty computation. Let the focal paper network be $G_{focal}$, with MeSH terms denoted by $t_i$, and let the historical network be $G_{past}$, constructed from historical papers $P$ and terms $T$. For a pair $(t_i, t_j)$, pairwise distances are computed as follows:

(a) **Term-Paper Co-occurrence**: A binary term–paper matrix is constructed from historical papers. The vector $v_i$ for term $t_i$ indicates its presence across historical papers. We then calculate the cosine distance between $v_i$ and $v_j$ to measure the pairwise distance:
$$Distance(t_i, t_j) = 1 - \frac{v_i \cdot v_j}{\|v_i\| \|v_j\|} \quad (1)$$

(b) **Term-Term Co-occurrence**: Each vector $v_i$ captures co-occurrence counts between $t_i$ and all other terms. The cosine distance between these vectors captures first- and second-order proximity.

(c) **Embedding**: We apply the LINE algorithm (Tang et al., 2015) to learn 200-dimensional embeddings for MeSH terms based on the historical network, then compute the cosine distance between embeddings, capturing first-order and second-order proximities in the historical network.

(d) **Geodesic Distance**: We compute the unweighted shortest path between $t_i$ and $t_j$ in $G_{past}$ using Dijkstra's algorithm (Dijkstra, 2017), denoted as $path_{i,j}$. Subsequently, pairwise distance is defined as
$$Distance(t_i, t_j) = 1 - \frac{1}{path_{i,j} + 1} \quad (2)$$

(e) **Geodesic Distance with Direct Connection Adjustment**: For directly connected node pairs, we use distance (b) derived from the term–term co-occurrence matrix as the value of $path_{i,j}$ in Equation (2) to differentiate varying degrees of novelty.

---
[3] https://developers.google.com/optimization

We denote these five methods as *Term–Paper*, *Term–Term*, *Embed*, *Geo_distance*, and *Geo_distance_weight*. **Figure 3** illustrates these approaches. Across all methods, pairwise distances lie in the interval (0,1]. Because none of the methods can compute distances for edges that include at least one new node, such edges are assigned a novelty value of 1. This is the maximum value attainable by any distance function above and captures the fact that a new concept should be considered far away from old concepts until connections are identified by researchers. An analysis of edge-type proportions shows that *new node new edges* constitute less than 1% of all edges after 1945 and only 0.14% by 2020, indicating that this treatment has negligible impact on paper-level novelty scores.

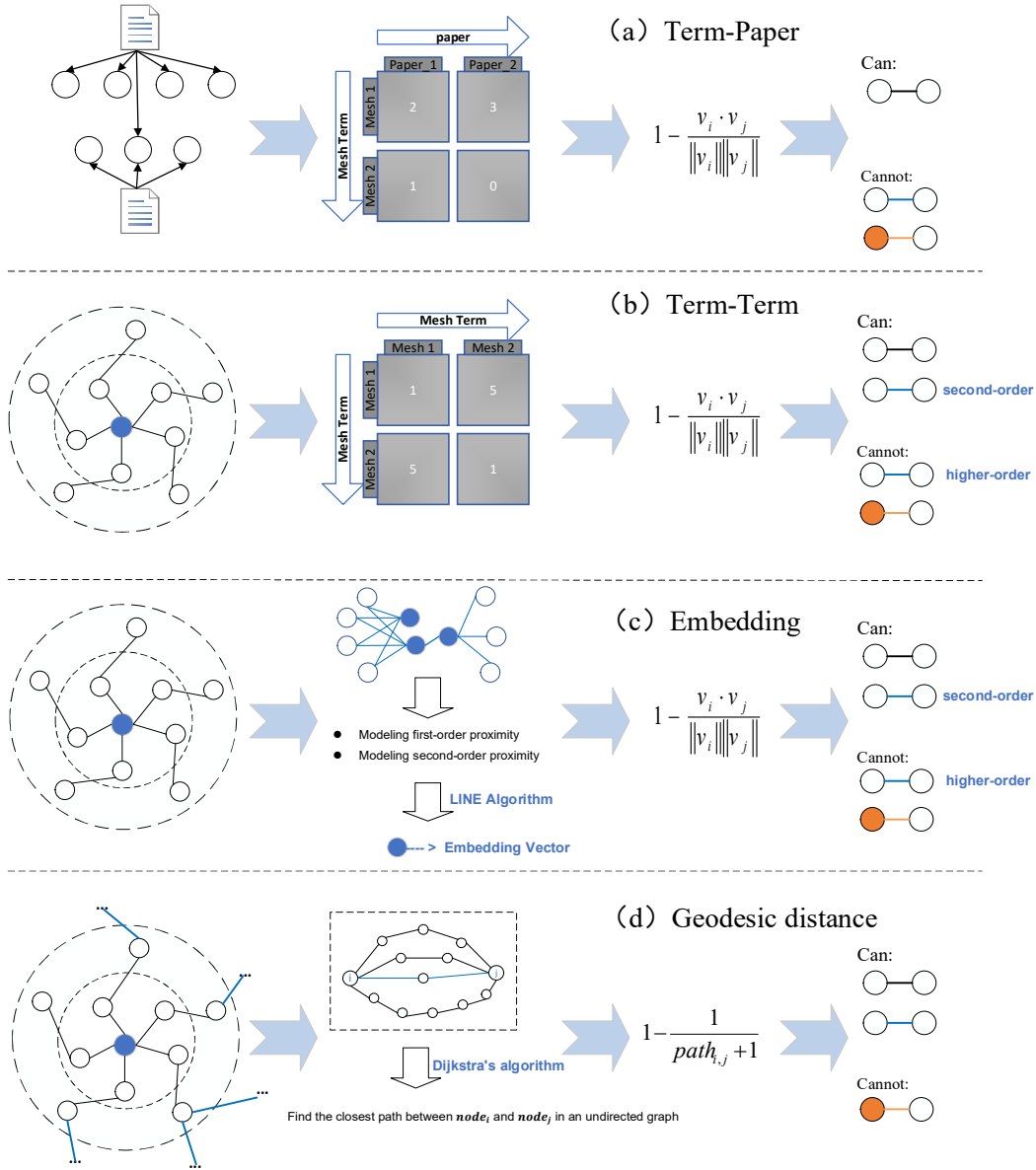

**Figure 2. Comparison of Pairwise Distance Computation Methods.** Illustration of four pairwise distance computation methods: (a) Term–Paper Co-occurrence, (b) Term–Term Co-occurrence, (c) Embedding, and (d) Geodesic Distance. The fifth method, (e) Geodesic Distance with Direct-Connection Adjustment, is derived from panel (d) and therefore not illustrated separately. Each panel depicts the historical network representation, the corresponding computational procedure, and the resulting pairwise distance. "Can" and "Cannot" indicate whether a method can differentiate varying levels of distance across specific edge types (e.g., old–old, old–new, and new–new connections) or instead assigns such edges a fixed novelty value of 1.

## 4.4 Result Validation

To validate CTD against human judgments of novelty, we construct two validation datasets: F1000Prime and Nobel Prize papers.

F1000Prime collects expert-recommended biomedical papers annotated with qualitative labels and is widely used in scientometric studies as an evaluation corpus (Waltman & Costas, 2014). Following Bornmann et al. (2019), we treat four labels—Interesting Hypothesis, New Finding, Novel Drug Target, and Technical Advance—as indicators of novelty. We identify 105,812 F1000Prime papers published between 2010 and 2020 and classify them as novel or non-novel accordingly. To address class imbalance, we perform matched sampling by publication year, journal, and subfield (based on OpenAlex classifications), yielding 8,902 matched pairs. We estimate Probit models with year and subfield fixed effects, using paper novelty (based on F1000Prime) as the dependent variable and CTD as the key independent variable.

Nobel Prize papers represent groundbreaking scientific contributions and thus provide a complementary validation benchmark. Using the dataset compiled by Li et al. (2019), we identify 208 Nobel Prize papers in our dataset. Each Nobel paper is matched with a control paper by publication year, journal, and subfield, resulting in 172 matched pairs. Probit regressions with fixed effects are used to assess whether CTD distinguishes Nobel-winning papers from matched non-Nobel-winning papers.

## 5. Results

### 5.1 Descriptive Analysis

Our dataset comprises 27,423,449 biomedical papers published between 1900 and 2020. **Figure 4** reports the annual publication volume and the average number of MeSH terms per paper. While publication volume exhibits a persistent upward trend, the average number of MeSH terms shows substantial fluctuation in early years. This arises primarily because terms were retroactively assigned to publications predating the MeSH system's formal introduction in 1954. As the vocabulary matured and standardized, term assignment stabilized; in recent decades, the average has converged to approximately ten, indicating a stable representation of knowledge content.

We further examine the composition of edge types in focal paper networks (**Figure 5**). A pronounced structural shift occurs around 1945, after which the proportion of edges involving new nodes declines sharply. This reflects the institutionalization of the MeSH system: once the core vocabulary was established, the introduction of entirely new terms became rare, meaning nearly all knowledge units in post-1945 papers already exist in the historical network.

This empirical pattern has two implications. First, the low prevalence of new-node edges implies that assigning them a fixed novelty value of 1 has a negligible effect on paper-level scores. Second, the dominance of old–old edges indicates that most papers rely primarily on established knowledge combinations, with genuinely novel pairwise combinations constituting only a small fraction of connections.

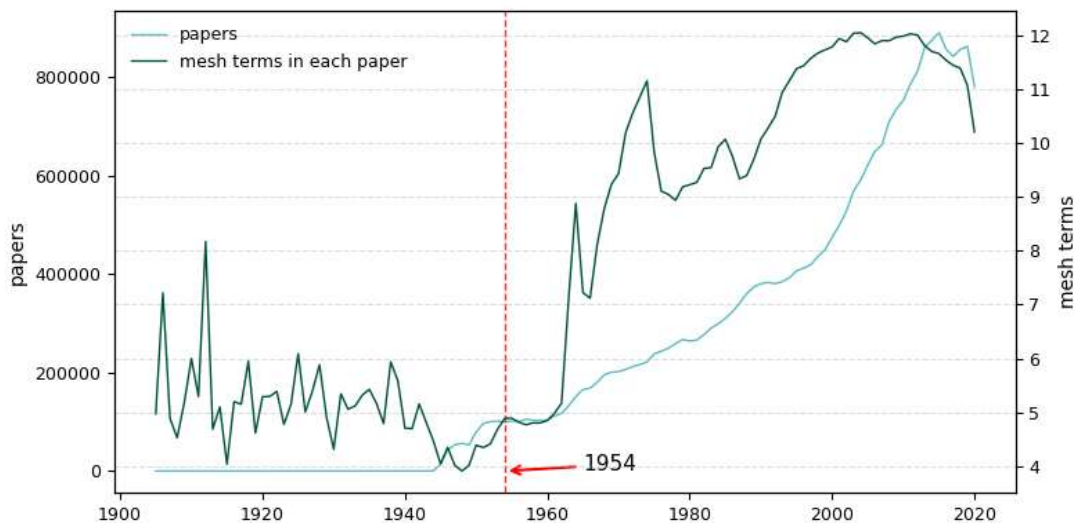

Figure 3. Annual Publication Volume and Average Number of MeSH Terms per Paper

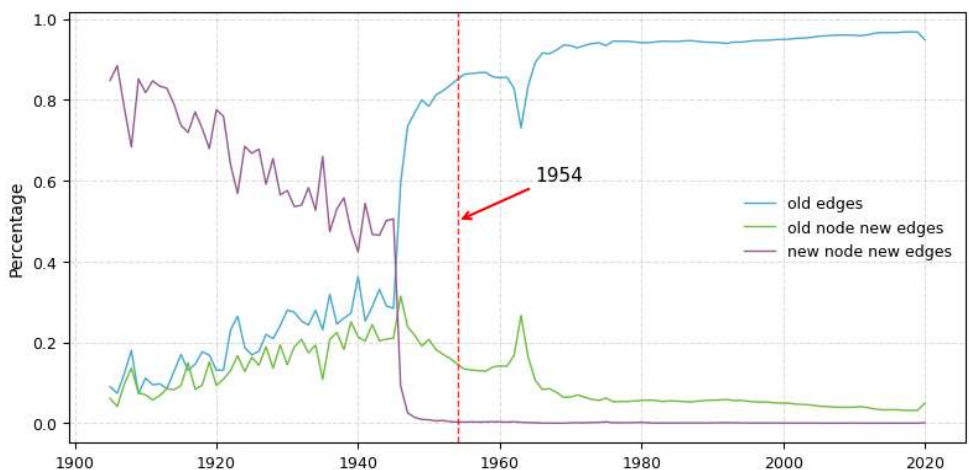

Figure 4. Annual Proportions of Edge Types in Focal Paper's Subgraphs

## 5.2 Regression Results

To better capture the effectiveness of the CTD method, we benchmark it against two traditional pairwise-based approaches: mean-based and 90th-percentile-based measurement. These approaches define a paper's novelty as either the average or upper-tail novelty across all knowledge pairs in the focal paper. We further compare CTD with a recent high-performing method that measures novelty by introducing new words and word combinations in scientific papers (Arts et al., 2025). This method has been shown to distinguish novel from non-novel papers across multiple validation settings.

### 5.2.1 Comparison with Pairwise Distance Indicators

We evaluate the performance of CTD using the F1000Prime and Nobel Prize datasets. **Table 2** reports regression results for the F1000Prime dataset. Panels A, B, and C correspond to CTD, mean-based aggregation, and 90th-percentile aggregation, respectively. Within each panel, results are shown for alternative methods of computing pairwise distances.

Across models, most novelty indicators exhibit significant positive associations with expert assessments, indicating that MeSH-term-based recombination measures are informative. Two exceptions occur under the mean-based aggregation, where the *Term–Paper* and *Embed* indicators yield negative coefficients. This pattern suggests that averaging pairwise novelty scores can obscure structural information and, in some cases, distort the novelty signal.

Comparisons of model fit provide clear evidence of the advantage of CTD. Pseudo-$R^2$ values are consistently higher under CTD than under either the mean or the 90th-percentile aggregation, indicating greater explanatory power. This advantage is reinforced by classification performance: for a given pairwise distance computation method, CTD systematically yields higher AUC values. For example, when pairwise novelty is computed using the Term–Term approach, the AUC under CTD reaches 60.58%, compared to 52.43% and 52.24% under the mean and 90th-percentile aggregations, respectively. Across alternative pairwise distance measures, CTD improves AUC by approximately 4–8 percentage points, highlighting the benefit of incorporating the global configuration of knowledge units rather than relying on dyadic summaries.

Performance also varies across pairwise distance methods. Under CTD, *Term–Term* and *Geo_distance_weight* perform best, as both capture novelty between indirectly connected knowledge units, with the latter further differentiating longer historical paths. Under mean and 90th-percentile aggregation, *Geo_distance* and *Term–Paper* perform best, respectively. By contrast, the *Embed* method consistently yields the weakest results, suggesting that embedding-based similarity alone may inadequately capture novelty-relevant distinctions.

**Table 3** reports corresponding results for the Nobel Prize dataset. Under traditional aggregation strategies (mean and 90th percentile), none of the novelty measures exhibit significant coefficients, and both Pseudo-$R^2$ and AUC values remain close to random classification. This finding suggests that conventional pairwise novelty metrics fail to distinguish Nobel Prize–winning papers from other publications within the same journal. In contrast, when novelty is measured using CTD, all novelty measures, except Embed, show significant positive associations with Nobel Prize status, and model AUC values increase substantially. Among the pairwise measures, *Geo_distance_weight* achieves the strongest performance, underscoring the importance of accounting for higher-order relational structure in the historical knowledge network when assessing scientific novelty.

In addition, our analysis randomly selected a non-Nobel paper for each Nobel Prize paper from publications with the same year, journal, and research topic. Because the number of Nobel Prize papers is small, this matching process may introduce variance in the results. To assess robustness, we therefore repeated the matching and estimation procedure ten times, drawing a different set of control papers in each run, and aggregated results across repetitions. **Figure 6** summarizes the variation in coefficient significance and model performance across novelty measurement methods. Except for the *Embed-based* pairwise measure, novelty indicators derived from CTD consistently exhibit significant positive associations with Nobel Prize status. Across repetitions, average p-values for CTD-based indicators remain below 0.05 for all pairwise distance computation methods. Moreover, AUC comparisons show that CTD not only achieves higher discriminative performance than traditional aggregation-based approaches but also exhibits greater robustness to control-sample selection. In contrast, mean- and percentile-based novelty measures display substantially larger variability in both significance and predictive accuracy across repeated samples.

Table 2. Prediction of F1000Prime Novel Papers Using Alternative Novelty Measures

| | F1000Prime Novel (vs. not) | | | | |
|---|---|---|---|---|---|
| | Probit | | | | |
| **Panel A: Cognitive Traversal Distance** | | | | | |
| | (1) | (2) | (3) | (4) | (5) |
| Term-Paper | 0.045*** | | | | |
| | (0.002) | | | | |
| Term-Term | | 0.291*** | | | |
| | | (0.013) | | | |
| Embed | | | 0.242*** | | |
| | | | (0.021) | | |
| Geo_distance | | | | 0.074*** | |
| | | | | (0.004) | |
| Geo_distance_weight | | | | | 0.370*** |
| | | | | | (0.016) |
| Pseudo-$R^2$ | 0.0147 | 0.0224 | 0.0058 | 0.0161 | 0.0236 |
| AUC | 0.5837 | 0.6058 | 0.5610 | 0.5868 | 0.6077 |
| **Panel B: Mean Pairwise Distance** | | | | | |
| | (6) | (7) | (8) | (9) | (10) |
| Term-Paper | -0.660** | | | | |
| | (0.242) | | | | |
| Term-Term | | 0.692*** | | | |
| | | (0.126) | | | |
| Embed | | | -0.019 | | |
| | | | (0.141) | | |
| Geo_distance | | | | 4.410*** | |
| | | | | (0.814) | |
| Geo_distance_weight | | | | | 1.272*** |
| | | | | | (0.216) |
| Pseudo-$R^2$ | 0.0003 | 0.0012 | 0.0000 | 0.0013 | 0.0014 |
| AUC | 0.5128 | 0.5243 | 0.4960 | 0.5452 | 0.5263 |
| **Panel C: 90th Percentile Pairwise Distance** | | | | | |
| | (11) | (12) | (13) | (14) | (15) |
| Term-Paper | 20.02*** | | | | |
| | (1.949) | | | | |
| Term-Term | | 0.434*** | | | |
| | | (0.078) | | | |
| Embed | | | 0.160** | | |
| | | | (0.052) | | |
| Geo_distance | | | | 0.436** | |
| | | | | (0.158) | |
| Geo_distance_weight | | | | | 0.814*** |
| | | | | | (0.159) |
| Pseudo-$R^2$ | 0.0047 | 0.0013 | 0.0004 | 0.0003 | 0.0011 |
| AUC | 0.5388 | 0.5224 | 0.5020 | 0.5058 | 0.5232 |

Notes: N=17,804(balanced sample). *$p<0.05$, **$p<0.01$, ***$p<0.001$

**Table 3. Prediction of Nobel Prize Papers Using Alternative Novelty Measures**

|  | Nobel Prize (vs. not) Probit | | | | |
|---|---|---|---|---|---|
| **Panel A: Cognitive Traversal Distance** | | | | | |
|  | (1) | (2) | (3) | (4) | (5) |
| Term-Paper | 0.085** (0.028) | | | | |
| Term-Term |  | 0.192* (0.089) | | | |
| Embed |  |  | 0.152 (0.153) | | |
| Geo_distance |  |  |  | 0.147** (0.048) | |
| Geo_distance_weight |  |  |  |  | 0.291* (0.123) |
| Pseudo-$R^2$ | 0.0202 | 0.0103 | 0.0020 | 0.0209 | 0.0125 |
| AUC | 0.5856 | 0.5747 | 0.5232 | 0.5906 | 0.5799 |
| **Panel B: Mean Pairwise Distance** | | | | | |
|  | (6) | (7) | (8) | (9) | (10) |
| Term-Paper | 0.410 (2.623) | | | | |
| Term-Term |  | -0.805 (0.788) | | | |
| Embed |  |  | 0.391 (0.682) | | |
| Geo_distance |  |  |  | 1.169 (1.671) | |
| Geo_distance_weight |  |  |  |  | -1.107 (1.263) |
| Pseudo-$R^2$ | 0.0001 | 0.0021 | 0.0007 | 0.0010 | 0.0016 |
| AUC | 0.4974 | 0.5061 | 0.5047 | 0.5178 | 0.5112 |
| **Panel C: 90th Percentile Pairwise Distance** | | | | | |
|  | (11) | (12) | (13) | (14) | (15) |
| Term-Paper | 1.432 (3.345) | | | | |
| Term-Term |  | -0.354 (0.605) | | | |
| Embed |  |  | 0.244 (0.324) | | |
| Geo_distance |  |  |  | 0.749 (0.681) | |
| Geo_distance_weight |  |  |  |  | -0.791 (0.898) |
| Pseudo-$R^2$ | 0.0004 | 0.0007 | 0.0012 | 0.0026 | 0.0015 |
| AUC | 0.4791 | 0.4905 | 0.5333 | 0.5363 | 0.4891 |

Note: N=348(balanced sample). *$p<0.05$, **$p<0.01$, ***$p<0.001$

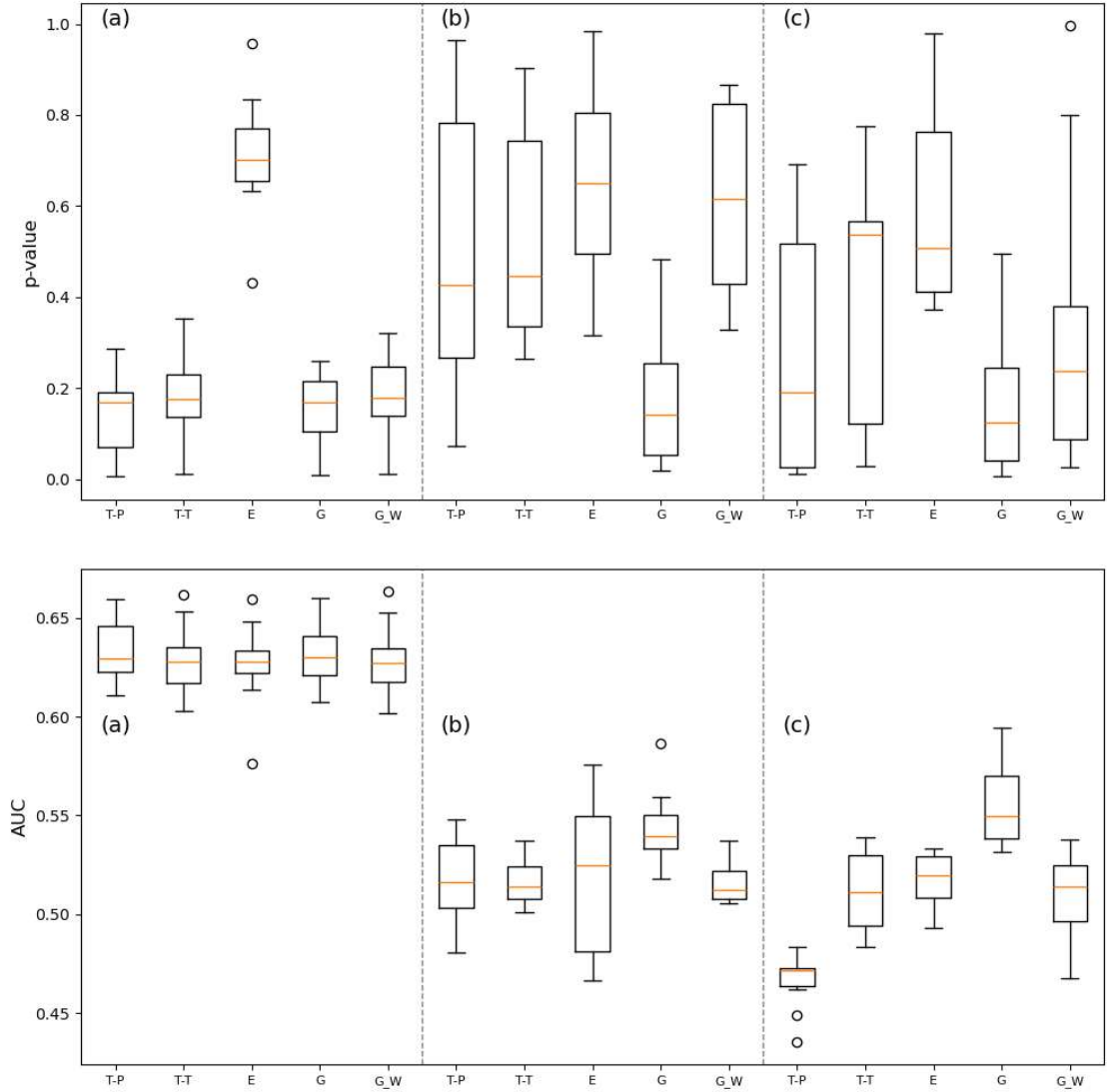

**Figure 5: Distribution of Coefficient P-values and AUC Across Repeated Nobel Prize Validation Runs**. Distribution of p-values (Upper panel) for the novelty indicator coefficients and AUC (Lower panel), across ten repeated matching and estimation runs using the Nobel Prize dataset. Subfigures (a), (b), and (c) correspond to three paper-level novelty measures: CTD, mean pairwise distance, and 90th-percentile pairwise distance, respectively. T–T, T–P, E, G, and G_W denote the five pairwise distance computation methods: Term–Term, Term–Paper, Embed, Geo_distance, and Geo_distance_weight, respectively.

## 5.2.2 Comparison with Recent High-Performing Text-based Novelty Measures

We benchmark CTD against four text-based novelty indicators (*New_word*, *New_phrase*, *New_word_comb*, *New_phrase_comb*) from Arts et al. (2025)[4]. Using the matching procedures described in Section 4.4, we constructed evaluation datasets comprising 8,636 pairs for F1000Prime and 142 pairs for Nobel Prize papers. We estimated Probit models including both CTD-based and text-based indicators, incorporating fixed effects and controls (e.g., text length, abstract availability) to replicate the experimental conditions in Arts et al. (2025).

Regression results are reported in **Tables 4 and 5**. For the F1000Prime dataset (**Table 4**), novelty measures based on *New_word_comb* and *New_phrase_comb* do not significantly distinguish novel from non-novel papers. Although *New_word* and *New_phrase* exhibit significant positive coefficients, their associated pseudo-$R^2$ and AUC values are consistently lower than those achieved by any of the CTD-based indicators. These results suggest that CTD aligns more closely with expert evaluations of novelty in this setting.

Results for the Nobel Prize dataset (**Table 5**) present a more nuanced pattern. In this context, the indicators proposed by Arts et al. (2025) achieve higher AUC values than the CTD-based measures. Analyses of coefficient significance and predictive performance across ten repeated matching runs (**Figure 7**) further confirm this advantage. When distinguishing Nobel Prize–winning papers from other high-quality publications, novelty measures that directly capture the introduction of new noun phrases and phrase combinations perform particularly well.

We interpret this difference as reflecting the distinct nature of novelty captured by the two approaches. Nobel Prize–winning papers often introduce entirely new research concepts or terminology, whereas CTD is grounded in an expert-curated and relatively stable knowledge classification system (MeSH). As a result, although CTD continues to significantly differentiate Nobel Prize papers from matched controls, MeSH-based traversal may be less sensitive to the emergence of brand-new conceptual labels. This reflects an important limitation of the CTD measure. Accordingly, CTD should be interpreted as a measure of integrative novelty within the existing knowledge landscape, rather than as a comprehensive indicator of all forms of scientific novelty.

---

[4] Data repository from Arts et al. (2025): https://zenodo.org/records/13869486

**Table 4. Prediction of F1000Prime Novel Papers Using CTD and Text-Based Novelty Measures**

| | F1000Prime Novel (vs. not) | | | | | | | | |
|---|---|---|---|---|---|---|---|---|---|
| | Probit | | | | | | | | |
| | (1) | (2) | (3) | (4) | (5) | (6) | (7) | (8) | (9) |
| Term-Paper | 0.032*** | | | | | | | | |
| | (0.003) | | | | | | | | |
| Term-Term | | 0.207*** | | | | | | | |
| | | (0.014) | | | | | | | |
| Embed | | | 0.178*** | | | | | | |
| | | | (0.022) | | | | | | |
| Geo_distance | | | | 0.053*** | | | | | |
| | | | | (0.004) | | | | | |
| Geo_distance_weight | | | | | 0.264*** | | | | |
| | | | | | (0.018) | | | | |
| New_word | | | | | | 0.108** | | | |
| | | | | | | (0.0289) | | | |
| New_phrase | | | | | | | 0.063*** | | |
| | | | | | | | (0.0131) | | |
| New_word_comb | | | | | | | | 0.0004 | |
| | | | | | | | | (0.000) | |
| New_phrase_comb | | | | | | | | | -0.001 |
| | | | | | | | | | (0.000) |
| Pseudo-$R^2$ | 0.0449 | 0.0481 | 0.0410 | 0.0456 | 0.0487 | 0.0387 | 0.0392 | 0.0383 | 0.0383 |
| AUC | 0.6455 | 0.6526 | 0.6430 | 0.6454 | 0.6528 | 0.6398 | 0.6412 | 0.6378 | 0.6377 |

Note: N=17,272(balanced sample). *p<0.05, **p<0.01, ***p<0.001

**Table 5. Prediction of Nobel Prize Papers Using CTD and Text-Based Novelty Measures**

| | Nobel Prize (vs. not) | | | | | | | | |
|---|---|---|---|---|---|---|---|---|---|
| | Probit | | | | | | | | |
| | (1) | (2) | (3) | (4) | (5) | (6) | (7) | (8) | (9) |
| Term-Paper | 0.080** | | | | | | | | |
| | (0.030) | | | | | | | | |
| Term-Term | | 0.251* | | | | | | | |
| | | (0.101) | | | | | | | |
| Embed | | | 0.066 | | | | | | |
| | | | (0.168) | | | | | | |
| Geo_distance | | | | 0.132** | | | | | |
| | | | | (0.051) | | | | | |
| Geo_distance_weight | | | | | 0.334* | | | | |
| | | | | | (0.135) | | | | |
| New_word | | | | | | 0.066 | | | |
| | | | | | | (0.142) | | | |
| New_phrase | | | | | | | 0.004 | | |
| | | | | | | | (0.002) | | |
| New_word_comb | | | | | | | | 0.282*** | |
| | | | | | | | | (0.078) | |
| New_phrase_comb | | | | | | | | | 0.031*** |
| | | | | | | | | | (0.007) |
| | | | | | | | | | |
| Pseudo-$R^2$ | 0.0334 | 0.0305 | 0.0157 | 0.0329 | 0.0303 | 0.0129 | 0.0200 | 0.0500 | 0.0697 |
| AUC | 0.6110 | 0.6030 | 0.5763 | 0.6074 | 0.6019 | 0.5841 | 0.5956 | 0.6565 | 0.6651 |

Note: N=284(balanced sample). *p<0.05, **p<0.01, ***p<0.001

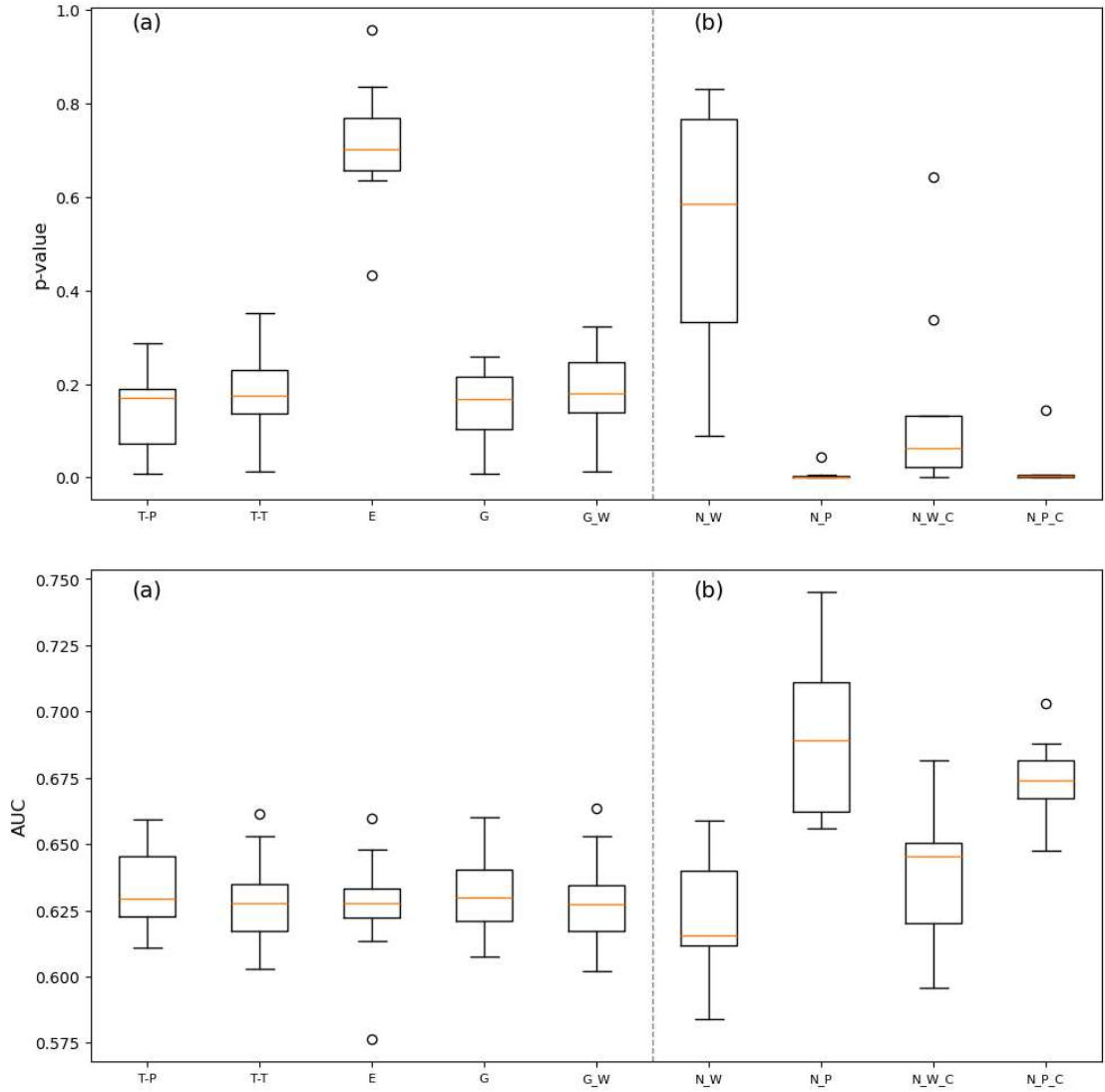

**Figure 7: Distribution of Coefficient P-values and AUC Across Repeated Nobel Prize Validation Runs, Comparing CTD and New-Word-Based Measures**. Distribution of p-values (Upper panel) for the novelty indicator coefficients and AUC (Lower panel), across ten repeated matching and estimation runs using the Nobel Prize dataset. Subfigures (a) and (b) correspond to CTD and new-word-based text measures. T–T, T–P, E, G, and G_W denote the five pairwise distance computation methods: *Term–Term*, *Term–Paper*, *Embed*, *Geo_distance*, and *Geo_distance_weight*, respectively. N_W, N_P, N_W_C, and N_P_C represent the four metrics introduced in Arts et al.'s study: *new_word*, *new_phrase*, *new_word_comb*, and *new_phrase_comb*, respectively.

## 5.3 Relationship between Paper Novelty and Citations

As a supplementary analysis, we examine how paper-level novelty relates to subsequent citations. Citation counts were collected for all papers in our dataset up to March 24, 2025. To reduce distortions associated with very early publications and incomplete citation windows, we restrict the analysis to papers published between 1980 and 2020, yielding a sample of 22,170,033 papers.

We estimate generalized negative binomial (GNB) regression models, which allow the dispersion of citation outcomes to vary with covariates. Control variables include the number of authors, institutions, and references, an indicator for international collaboration, the number of MeSH terms, and fixed effects for publication year and research field. Regression results are reported in **Table 6**. Across all models, the dispersion coefficients associated with novelty are positive, indicating that higher novelty is linked to greater variability in subsequent citation impact, consistent with the theoretical expectation (Wang et al., 2017).

We further examine whether novelty predicts the likelihood that a paper becomes highly cited, defined as being among the top 1% of cited papers within the same publication year and research field. Results from these models are shown in **Table 7**. Across all five CTD-based novelty indicators, higher novelty scores are associated with a significantly greater probability of becoming highly cited. In contrast, under mean- and percentile-based aggregation strategies, certain pairwise novelty indicators, i.e., *Geo_distance* and *Term–Paper*, exhibit negative associations with top-citation status, suggesting greater sensitivity to isolated dyadic relations.

Comparisons of model fit using AIC and BIC further indicate that CTD consistently provides superior explanatory performance relative to traditional aggregation-based novelty measures. Taken together, these findings suggest that traversal-based novelty captures aspects of scientific novelty that are not only conceptually meaningful ex ante, but also consequential for downstream scientific recognition.

**Table 6. Prediction of Citation Mean and Dispersion Using CTD**

|  | Citations | | | | |
|---|---|---|---|---|---|
|  | GNB | | | | |
|  | (1) | (2) | (3) | (4) | (5) |
| Mean | | | | | |
| Term-Paper | 0.019*** <br> (0.000) | | | | |
| Term-Term | | 0.016*** <br> (0.000) | | | |
| Embed | | | -0.012*** <br> (0.001) | | |
| Geo_distance | | | | 0.055*** <br> (0.000) | |
| Geo_distance_weight | | | | | 0.013*** <br> (0.001) |
| Dispersion | | | | | |
| Term-Paper | 0.049*** <br> (0.001) | | | | |
| Term-Term | | 0.027*** <br> (0.001) | | | |
| Embed | | | 0.100*** <br> (0.001) | | |
| Geo_distance | | | | 0.112*** <br> (0.001) | |
| Geo_distance_weight | | | | | 0.048*** <br> (0.001) |
| Pseudo-$R^2$ | 0.0342 | 0.0341 | 0.0342 | 0.0340 | 0.0341 |
| Log lik | -9441968 | -9442442 | -9441930 | -9439386 | -9442398 |

Note: N=22,170,033. $^*p<0.05$, $^{**}p<0.01$, $^{***}p<0.001$

## Table 7. Prediction of Top 1% Highly Cited Using CTD and Pairwise-Distances

|  | Top 1% Cited (vs. not) |  |  |  |  |
|---|---|---|---|---|---|
|  | Probit |  |  |  |  |

**Panel A: Cognitive Traversal Distance**

|  | (1) | (2) | (3) | (4) | (5) |
|---|---|---|---|---|---|
| Term-Paper | 0.040*** |  |  |  |  |
|  | (0.001) |  |  |  |  |
| Term-Term |  | 0.024*** |  |  |  |
|  |  | (0.002) |  |  |  |
| Embed |  |  | 0.037*** |  |  |
|  |  |  | (0.003) |  |  |
| Geo_distance |  |  |  | 0.092*** |  |
|  |  |  |  | (0.001) |  |
| Geo_distance_weight |  |  |  |  | 0.033*** |
|  |  |  |  |  | (0.002) |
| AIC | 2,208,421 | 2,211,991 | 2,212,046 | 2,207,841 | 2,211,999 |
| BIC | 2,209,495 | 2,213,065 | 2,213,120 | 2,208,914 | 2,213,073 |
| Pseudo-$R^2$ | 0.1125 | 0.1110 | 0.1110 | 0.1127 | 0.1110 |

**Panel B: Mean Pairwise Distance**

|  | (6) | (7) | (8) | (9) | (10) |
|---|---|---|---|---|---|
| Term-Paper | 0.543*** |  |  |  |  |
|  | (0.024) |  |  |  |  |
| Term-Term |  | 0.120*** |  |  |  |
|  |  | (0.010) |  |  |  |
| Embed |  |  | 0.089*** |  |  |
|  |  |  | (0.012) |  |  |
| Geo_distance |  |  |  | -0.327*** |  |
|  |  |  |  | (0.058) |  |
| Geo_distance_weight |  |  |  |  | 0.111*** |
|  |  |  |  |  | (0.018) |
| AIC | 2,211,723 | 2,212,121 | 2,212,206 | 2,212,232 | 2,212,229 |
| BIC | 2,212,797 | 2,213,194 | 2,213,280 | 2,213,305 | 2,213,303 |
| Pseudo-$R^2$ | 0.1111 | 0.1110 | 0.1109 | 0.1109 | 0.1109 |

**Panel C: 90th Percentile Pairwise Distance**

|  | (11) | (12) | (13) | (14) | (15) |
|---|---|---|---|---|---|
| Term-Paper | -3.936*** |  |  |  |  |
|  | (0.080) |  |  |  |  |
| Term-Term |  | 0.192*** |  |  |  |
|  |  | (0.007) |  |  |  |
| Embed |  |  | 0.077*** |  |  |
|  |  |  | (0.004) |  |  |
| Geo_distance |  |  |  | 0.066*** |  |
|  |  |  |  | (0.011) |  |
| Geo_distance_weight |  |  |  |  | 0.371*** |
|  |  |  |  |  | (0.013) |
| AIC | 2,210,872 | 2,211,350 | 2,211,961 | 2,212,230 | 2,211,337 |
| BIC | 2,211,946 | 2,212,424 | 2,213,034 | 2,213,304 | 2,212,411 |
| Pseudo-$R^2$ | 0.1115 | 0.1113 | 0.1110 | 0.1109 | 0.1113 |

## 5.4 Robustness Checks

We conducted a series of robustness checks to assess whether our results are sensitive to alternative sampling strategies, model specifications, and parameter choices.

First, we examined the robustness of our findings to alternative strategies for sample construction and control group matching. To assess sensitivity to sampling decisions, we re-analyzed the F1000Prime dataset using the full set of papers without subsampling, as well as by resampling the majority class without restricting the number of matched papers. We further evaluated three alternative matching schemes for both datasets by requiring control papers to share the same: (1) publication year only, (2) year and journal, and (3) year, journal, and research topic (using OpenAlex classifications). Across all specifications, CTD consistently outperformed traditional aggregation-based novelty measures in both coefficient significance and model fit.

Second, we assessed the potential influence of variation in the number of MeSH terms per paper. To do so, we normalized the CTD-based novelty score by the number of MeSH terms associated with each paper. Regression analyses based on this normalized measure continue to show that traversal-based novelty exhibits greater explanatory power than mean-based and 90th-percentile-based aggregation strategies. However, apart from the *Geo_distance* specification, the explanatory power of models using other pairwise novelty computation methods decreases modestly after normalization. This pattern suggests that part of the novelty signal captured by CTD reflects the scale of knowledge integration, which normalization partially attenuates.

Third, we varied the time window used to construct the historical knowledge network, reducing it from five years to three years. The main results remain substantively unchanged under this alternative specification, indicating that our findings are not driven by a particular choice of historical time window length.

Taken together, these robustness checks demonstrate that our results are stable across alternative sampling strategies, matching criteria, normalization schemes, and temporal specifications, reinforcing the reliability of the proposed traversal-based approach to measuring scientific novelty.

## 6. Discussion and Conclusion

This study advances the measurement of scientific novelty within the framework of combinatorial novelty by addressing a fundamental limitation of existing approaches: reliance on aggregating pairwise distances between knowledge units to characterize paper-level novelty. Rather than treating novelty as an additive or extreme-value property of dyadic combinations, we conceptualize scientific novelty as a cognitive traversal across a structured knowledge landscape. We introduce Cognitive Traversal Distance, a holistic novelty measure defined as the length of the shortest traversal required to connect all knowledge units given their historical conceptual distances. CTD captures the minimal cognitive effort required to integrate multiple knowledge units into a coherent scientific contribution.

The core intuition underlying CTD is that novelty is not determined solely by the presence of unusual or distant knowledge pairs, but by the global configuration of how knowledge units relate to one another. This perspective is consistent with research in organizational and collective innovation, which emphasizes that innovativeness emerges from interaction, coordination, and structural relationships rather than from simple aggregation of individual components (von Hippel & von Krogh, 2003; Muthukrishna & Henrich, 2016; Li et al., 2025). Scientific papers can similarly be viewed as "knowledge organizations," in which concepts, methods, and ideas are connected through logical and semantic relationships (Bazerman, 1988). Our empirical findings support this view: across expert-evaluated benchmarks, CTD aligns more closely with novelty assessments than conventional mean- or percentile-based aggregation of pairwise distances.

Importantly, our comparison with recent text-based novelty measures delineates a clear scope condition for CTD. While traversal-based novelty performs well in distinguishing expert-identified novel papers, particularly in the F1000Prime setting, it is less sensitive to novelty driven by the introduction of entirely new conceptual labels or terminology, as often characteristic of Nobel Prize–winning research. This limitation is intrinsic to CTD's reliance on expert-curated and relatively stable knowledge taxonomies such as MeSH. As a result, CTD should be interpreted as capturing integrative novelty within an established knowledge landscape, rather than all forms of scientific novelty. This distinction clarifies what CTD is designed to measure and helps reconcile differences in performance across evaluation benchmarks.

Our study has limitations that outline avenues for future research. First, given our focus on biomedical literature using MeSH terms, the generalizability of CTD to domains with different epistemic structures or classification systems remains to be tested. Second, the reliance on small, expert-curated validation datasets warrants future assessment on larger-scale peer-review corpora. Finally, while MeSH terms ensure consistency, their fixed structure may overlook emerging concepts. Future work should therefore explore flexible knowledge representations, such as keywords, or combine CTD with text-based indicators to capture broader dimensions of novelty.

In sum, this study demonstrates that modeling novelty as a global property of knowledge integration, rather than as an aggregation of independent pairwise distances, provides a theoretically grounded and empirically effective approach to measuring scientific novelty. By explicitly accounting for the structural relationships among knowledge units, Cognitive Traversal Distance offers a principled framework for understanding how new scientific contributions emerge within, and traverse, the existing knowledge landscape.

# Acknowledgments

The computational results were obtained using the Austrian Scientific Computing (ASC) infrastructure. This research was supported by the China Scholarship Council (CSC). Part of PWs work was supported by the Vienna Science and Technology Fund (WWTF) and the City of Vienna project StruDL (10.47379/ ICT22059).